# Clustering in Redshift Space: Linear Theory

S. Zaroubi[1,2] and Y. Hoffman[2]

(1) Astronomy Department and Center for Particle Astrophysics, University of California, Berkeley, CA 94720, U.S.A.

(2) Racah Institute of Physics, The Hebrew University, Jerusalem 91904, Israel

## ABSTRACT

The clustering in redshift space is studied here to first order within the framework of gravitational instability. The distortion introduced by the peculiar velocities of galaxies results in anisotropy in the galaxy distribution and mode-mode coupling when analyzed in Fourier space. An exact linear calculation of the full covariance matrix in both the real and Fourier space is presented here. The explicit dependence on $\Omega_0$ and the biasing parameter is calculated and its potential use as a probe of these parameters is analyzed. It is shown that Kaiser's formalism can be applied only to a data set that subtends a small solid angle on the sky, and therefore cannot be used in the case of all sky surveys. The covariance matrix in the real space is calculated explicitly for $CDM$ model, where the behavior along and perpendicular to the line of sight is shown.

Subject Headings *galaxies: clustering; cosmology: large scale structure of universe*

---

[1,2] E-mail: saleem@pac1.berkeley.edu
[2] E-mail: hoffman@vms.huji.ac.il

# I. INTRODUCTION

One of the basic tools of observational cosmology is the mapping of the galaxy distribution in redshift space, where the observed redshift is used rather than physical distances. Indeed, redshift surveys have played a major role in shaping our understanding of the large scale structure of the universe. Given the tremendous difficulties in obtaining true distances of galaxies, redshift surveys will remain the major method of studying the large scale distribution of galaxies. Various groups have determined the statistical properties of the galaxy distribution directly from galaxy redshift catalogs. In particular, the power spectrum has been evaluated from the galaxy redshift distribution (Boumgart and Fry 1991, Park, Gott and da Costa 1992, Fisher *et al.* 1993, Feldman, Kaiser and Peacock, 1993).

The most basic analysis of the large scale clustering of galaxies is that of the power spectrum estimation, or equivalently its Fourier conjugate the 2-point correlation function (*cf.* Peebles 1980). From the theoretical point of view the power spectrum can be easily calculated in a variety of models and thus the observationally estimated power spectrum can be used to yield significant information on the nature of the early universe and the process of structure formation. However, theoretical calculations usually give the real space power spectrum while observationally it is determined in redshift space. Now, the formation of structure out of an expanding Friedmann universe involves departure from homogeneity, and therefore by the continuity equation peculiar velocities arise. This leads to the displacement of galaxies along the line of sight and introduces anisotropy to the apparent clustering of galaxies. On small scales, corresponding to nonlinear perturbations, virial velocities in collapsed clusters leads to the so-called 'finger of god' effect (Peebles 1980). This results in the elongation of the structure along the line of sight in redshift space. On larger scales where deviations from the Friedmann universe are in the linear regime, the infall to overdense regions compresses the structures along the line of sight. The relation between the power spectrum in real and redshift space is model dependent and can be solved only within the framework of an assumed model.

At the present paper we shall focus on the theoretical calculation of the redshift space power spectrum in the linear regime. This was first calculated by Kaiser (1987) who used the single mode approximation to find a simple relation between the redshift and real space power spectra. This relation depends on the density parameter $\Omega_0$ and the biasing parameter $b$ (see discussion below), and thus can be used to find these important parameters. Kaiser's simple relation is rigorously correct in the case of a single Fourier mode, but as is shown in the present *paper* it is not valid in the general case of a multi-mode field which subtends a wide solid angle over the observer's sky. Nevertheless, Kaiser's formalism has been widely used by many workers in the field to reconstruct the real space power spectrum from the observed one and to set constraints on $\Omega_0$ and $b$. Therefore it cannot be applied to analyze the (almost) full sky redshift surveys such as the IRAS $1.2Jy$ redshift survey (Fisher *et al.* 1993) and the IRAS QDOT redshift survey (Feldman, Kaiser and Peacock 1993). This limitation has escaped the attention of many workers in the field and has been used in analyzing full sky surveys. The major aim of the present *paper* is to clarify this important point and to present a rigorous calculation of the correlation function and the power spectrum in redshift space. Here we shall focus only on the theoretical aspects of this calculation and in a forthcoming paper this will be applied to the IRAS data base (Bistolas, Zaroubi and Hoffman, 1994). The exact treatment of selection effects and shot noise will be given at that paper. The main results reported here are, first, that the anisotropy introduced by the redshift manifests itself by an off-diagonal coupling of the Fourier modes, where the Fourier space is the conjugate to the redshift space and the ratio of the power spectra of the two spaces is $k$-dependent. Thus, the power spectrum is not the Fourier transform of the two-point correlation function. Here, the term power spectrum is used in the more general sense of the covariance matrix in the Fourier space. The second result is that the correlation between any two points in the redshift configuration space depends not only on the magnitude of the displacement vector which connects them but also on its orientation and distance from the observer.

The structure of the *paper* is as follows. An exact (to first order) calculation of the Fourier modes covariance matrix in redshift space is presented in §II. The Kaiser's formalism is recovered here in the special case where the data subtends a small solid angle on the sky (§III). The covariance matrix in the redshift configuration space is presented and calculated explicitly for special cases in §IV. A general discussion is given in §V.

## II. LINEAR THEORY IN REDSHIFT SPACE

The coordinate transformation from **r**-space (real space) to **s**-space (redshift space) is given by:

$$\mathbf{s}(\mathbf{r}) = \mathbf{r}[1 + \frac{U(\mathbf{r}) - U(0)}{r}] \tag{1}$$

Here $U(\mathbf{r})$ and $U(0)$ are the projection of the peculiar velocity at the point **r** and at the position of the observer, respectively, on the radius vector connecting these two points. Distances are measured here in velocity units. The relation between the densities in the two spaces are given through the Jacobian,

$$\rho_s = \rho_r ||\frac{d^3 s}{d^3 r}||^{-1} = \rho_r [1 + \frac{U(\mathbf{r}) - U(0)}{r}]^{-2} [1 + \frac{\partial U(\mathbf{r})}{\partial r}]^{-1}. \tag{2}$$

The derivation here assumes that the selection effects are given by the selection function $\phi(r)$ and the measured density in redshift space is weighted with $\phi^{-1}(s)$. To first order (in perturbation theory) the fractional overdensities in **s**- and **r**-space are related by:

$$\delta_s(\mathbf{r}) = \delta_r(\mathbf{r}) - \left(2 + \frac{d(ln\phi)}{d(lnr)}\right) \frac{U(\mathbf{r}) - U(0)}{r} - \frac{\partial U(\mathbf{r})}{\partial r} \tag{3}$$

Note that in the second term of Eq. 3 $\left(U(\mathbf{r}) - U(0)\right)$ is of the order of the *rms* peculiar velocities and it is distance independent. This second term scales as the inverse of the depth of the survey and is usually neglected as compared to the other terms (Kaiser 1987). However, in our patch of the universe velocities are of the order of several $10^3 km/sec$ which is not negligible on scales of $\sim 10^4 km/sec$. Therefore, our treatment will in general include this term unless it is stated otherwise.

The relevant Fourier transformed fields are:

$$\delta_s(\mathbf{r}) = \int \exp(i\mathbf{r} \cdot \mathbf{k})\delta_s(\mathbf{k})\mathrm{d}^3 k \tag{4a}$$

$$\delta_r(\mathbf{r}) = \int \exp(i\mathbf{r} \cdot \mathbf{k})\delta_r(\mathbf{k})\mathrm{d}^3 k \tag{4b}$$

From linear theory (Peebles, 1980) one finds

$$U = -if(\Omega_0) \int \exp(i\mathbf{r} \cdot \mathbf{k}) \frac{\mathbf{k} \cdot \mathbf{r}}{kr} \delta_r(\mathbf{k})\mathrm{d}^3 k$$

$$\frac{\partial U}{\partial r} = -f(\Omega_0) \int \exp(i\mathbf{r} \cdot \mathbf{k})(\frac{\mathbf{k} \cdot \mathbf{r}}{kr})^2 \delta_r(\mathbf{k})\mathrm{d}^3 k, \tag{5}$$

where $f(\Omega_0) \sim \Omega_0^{0.6}$. In the case of linear biasing $f(\Omega_0)$ is to be replaced by $\beta = f(\Omega_0)/b$. In the inverse $k$-space Eq. 3 is written as:

$$\delta_s(\mathbf{k}) = \delta_r(\mathbf{k}) + V\, I_0(\mathbf{k}) + f(\Omega_0) \int d^3 k' \delta_r(\mathbf{k}') \left( I(\mathbf{k}', \mathbf{k}) - \frac{iI_2(\mathbf{k}, \mathbf{k}')}{k'} \right) \tag{6}$$

where $I_0(\mathbf{k})$ and the coupling kernels $I(\mathbf{k}', \mathbf{k}), I_1(\mathbf{k}', \mathbf{k})$ and $I_2(\mathbf{k}', \mathbf{k})$ are defined as follows:

$$I_0(\mathbf{k}) = \frac{1}{(2\pi)^3} \int \exp(-i\mathbf{k} \cdot \mathbf{r}) \left( 2 + \frac{d(ln\phi)}{d(lnr)} \right) \frac{\mathbf{k} \cdot \mathbf{r}}{kr} d^3 r \tag{7a}$$

$$I(\mathbf{k}', \mathbf{k}) = \frac{1}{(2\pi)^3} \int d^3 r \exp(i\mathbf{r} \cdot (\mathbf{k}' - \mathbf{k}))(\frac{\mathbf{k}' \cdot \mathbf{r}}{k'r})^2 \tag{7b}$$

$$I_1(\mathbf{k}', \mathbf{k}) = \delta_D(\mathbf{k} - \mathbf{k}') = \frac{1}{(2\pi)^3} \int \exp(i(\mathbf{k}' - \mathbf{k}) \cdot \mathbf{r}) d^3 r \tag{7c}$$

$$I_2(\mathbf{k}', \mathbf{k}) = \frac{1}{(2\pi)^3} \int \exp(i(\mathbf{k}' - \mathbf{k}) \cdot \mathbf{r}) \left( 2 + \frac{d(ln\phi)}{d(lnr)} \right) \frac{\mathbf{k}' \cdot \mathbf{r}}{k'r} d^3 r \tag{7d}$$

The crucial new result obtained here is that the transformation to redshift space variable, $\delta_s$, introduces mode-mode coupling . Now, in the standard model of cosmology the perturbation field is assumed to be Gaussian, and for a homogeneous and isotropic field, this implies that the covariance matrix of the real space density field is diagonal,

$$< \delta_r(\mathbf{k})\delta_r^*(\mathbf{k}') > = P(k)\delta_D(\mathbf{k} - \mathbf{k}') \tag{8}$$

and the covariance matrix is given by:

$$\begin{aligned}
< \delta_s(\mathbf{k}_1)\delta_s^*(\mathbf{k}_2) > = &\int d^3k'_1 P(\mathbf{k'}_1) \\
&\times \left[ I_1(\mathbf{k'}_1, \mathbf{k}_1) - \frac{if(\Omega_0)}{k'_1} I_2(\mathbf{k'}_1, \mathbf{k}_1) + f(\Omega_0) I(\mathbf{k'}_1, \mathbf{k}_1) \right] \\
&\times \left[ I_1^*(\mathbf{k'}_1, \mathbf{k}_2) + \frac{if(\Omega_0)}{k'_1} I_2^*(\mathbf{k'}_1, \mathbf{k}_2) + f(\Omega_0) I^*(\mathbf{k'}_1, \mathbf{k}_2) \right] \\
&+ V^2 I_0(\mathbf{k}_1) I_0^*(\mathbf{k}_2)
\end{aligned} \quad (9)$$

Neglecting the second term in equation (3) the covariance matrix thus obtained is,

$$\begin{aligned}
< \delta_s(\mathbf{k})\delta_s^*(\mathbf{k'}) > = &P(k)\delta_D(\mathbf{k}-\mathbf{k'}) \\
&+ f(\Omega_0)P(k')I(\mathbf{k'},\mathbf{k}) + f(\Omega_0)P(k)I^*(\mathbf{k},\mathbf{k'}) \\
&+ f^2(\Omega_0) \int d^3k'' P(k'') I(\mathbf{k''},\mathbf{k}) I^*(\mathbf{k''},\mathbf{k'})
\end{aligned} \quad (10)$$

which still includes a mode-mode coupling and it is much more complicated than the simple formula commonly used.

The transformation to redshift space changes the real space perturbation field in two fundamental ways. The first and obvious one is the anisotropy introduced by the line-of-sight peculiar velocity. This was recognized by Kaiser (1987) and in all subsequent works. The second and more subtle effect is the mode-mode coupling in the Fourier space representation. The source of this coupling is the geometric nature of the real − redshift space transformation which depends on the $\frac{\mathbf{k}\cdot\mathbf{r}}{kr}$ term. Many of the common properties of the real space perturbation field do not hold in redshift space. One such basic relation often used is that the real space 2-point correlation function and the power spectrum form a Fourier conjugate pair. This relation breaks down in redshift space as is manifested by Eq. 9 which shows that the statistical properties of the covariance matrix (i.e. the power spectrum) only. In any case, there is no simple $k$-independent relation between the real and redshift space power spectra. This is clearly manifested by the work of Fisher *et al.* (1993), which generated mock IRAS catalogs from cold dark matter model (CDM; Blumenthal *et al.* 1984) N-body simulations and compared the power spectra evaluated in the two spaces. Figs. 3b and 3c of Fisher *et al.* clearly shows a $k$ dependence of this relation, and in particular that this is inconsistent with Kaiser's relation. This result was also confirmed by Gramann, Cen and Bahcall (1993), who have analyzed N-body simulations in redshift and real space. They calculated the power spectrum in both spaces and found that the ratio of these is $k$ dependent. This is manifested by Fig. 2a of Gramann, Cen and Bahcall (1993). Last point to be made here concerns a property which holds also in redshift space. The redshift transformation is linear and therefore it preserves the Gaussian nature of the random field. Now, the $\delta_s(\mathbf{s})$ field is not isotropic and its ($k$-space) covariance matrix is not diagonal but its statistical properties can be analyzed within the standard framework of Gaussian random fields.

### III. COMPARISON WITH KAISER'S FORMALISM

The single mode approach of Kaiser (1987) is easily recovered from our more general formalism, we apply his assumptions to equation (6), namely by neglecting $I_0$ and $I_2$, and substituting

$$\delta_r(\mathbf{k}) = \delta_0 \delta_D(\mathbf{k} - \mathbf{k}_0) \tag{11}$$

we obtain Kaiser's result,

$$\delta_s(\mathbf{k}_0) = \delta_0(\mathbf{k}_0) + f(\Omega_0) \int d^3 r \left(\frac{\mathbf{k}_0 \cdot \mathbf{r}}{k_0 r}\right)^2 \delta_0(\mathbf{k}_0) \tag{12}$$

Expressing Eq. 12 in Kaiser's notation we define

$$\mu = \frac{\mathbf{k}_0 \cdot \mathbf{r}}{k_0 r} \tag{13}$$

to obtain:

$$\delta_s(\mathbf{k}_0) = \delta_0(\mathbf{k}_0)(1 + f(\Omega_0)\overline{\mu^2}) \tag{14}$$

Here the angle-averaged of $\mu^2$ is $\overline{\mu^2} = 1/3$. Kaiser's basic relation is

$$\delta_s(\mathbf{k}_0) = \delta_0(\mathbf{k}_0)(1 + f(\Omega_0)\mu^2), \tag{15}$$

without taking the angle-average. However, this expression is mathematically ill defined, as any **r** dependent quantity does not exist in $k$ space. Only after the angle- average is taken, namely integrating over real space, Kaiser's results is properly defined and is equal to the expression obtained here. In this sense Kaiser's result has been used by workers in the field (*eg.* Fisher *et al.* 1993, Feldman, Kaiser and Peacock, 1993), without realizing that it is rigorously correct only in the case of a single mode.

Kaiser's formalism is useful and accurate in the case where the data is confined within a volume which extends a small solid angle relative to the observer. In this case the $\mu$ is redefined to be:

$$\mu = \frac{\mathbf{k_0} \cdot (\mathbf{R} + \mathbf{r})}{\mathbf{k_0}\sqrt{\mathbf{r^2} + \mathbf{R^2}}} \quad (16)$$

Here **R** defines the center of that volume, in some general sense. Now, in the above limit of small solid angle, $R \gg r$, and

$$\mu \sim \frac{\mathbf{k}_0 \cdot \mathbf{R}}{k_0 R}. \quad (17)$$

Substituting this approximation of $\mu$ in Eqs. 5-10, the mode-mode coupling that is introduced by $\mu$ breaks down and Kaiser's results are fully recovered, namely Eq. 15 and

$$P_s(k) = P(k)(1 + f(\Omega_0)\mu^2)^2. \quad (18)$$

Indeed, this is the case studied by Cole, Fisher and Weinberg (1993), who analyzed redshift distortions in N-body simulations and confirmed Kaiser's formalism in the limit of samples that subtends a small solid angle.

The exact correlation matrix (Eq. 10) has a complicated mode-mode coupling, which is introduced by the $I(\mathbf{k}, \mathbf{k}')$ coupling term. It can be shown that for the diagonal terms in the limit of small wave number ($\mathbf{k} = \mathbf{k}' \to 0$) one obtains:

$$<|\delta_s(\mathbf{k})|^2> \sim P(\mathbf{k})\left(1 + \frac{2}{3}f(\Omega_0)\right) \quad (19)$$

Note that this differs from Kaiser's formula in not having the quadratic $f(\Omega_0)^2$ term, which for $f(\Omega_0) = 1.0$ amounts for a $\sim 12\%$ difference. It is interesting that the asymptotic relation found here is the same result quoted by Kaiser as the one derived from Peebles' (1980) formalism.

## IV. CORRELATION FUNCTION IN REDSHIFT SPACE

As discussed before the transformation to redshift space introduces anisotropy and mode–mode coupling in Fourier space representation. This complicates the picture in Fourier space and motivates us to calculate the correlation matrix in the redshift configuration space. The correlation matrix is expected to have three degrees of freedom ($d.o.f$), the first is the distance of one of the points from the observer and the other two are, the line of sight ($l.o.s.$) component of the displacement vector which connect the two points and the component perpendicular to it. The correlation function in redshift space for the points $\mathbf{r_1}$ and $\mathbf{r_2}$ can be calculated directly from Eq. 3. Let $\mathbf{r} = \mathbf{r_2} - \mathbf{r_1}$, $r = |\mathbf{r}|$ and $\hat{\mathbf{r}} = \frac{\mathbf{r}}{r}$, the correlation matrix is given by,

$$<\delta_s(\mathbf{s_1})\,\delta_s(\mathbf{s_2})> = \xi(r) - \frac{\Upsilon(r_2)}{r_2}\hat{\mathbf{r}}_2 \cdot \boldsymbol{\eta}(\mathbf{r}) + \frac{\Upsilon(r_1)}{r_1}\hat{\mathbf{r}}_1 \cdot \boldsymbol{\eta}(\mathbf{r}) - \hat{\mathbf{r}}_2 \cdot \partial\boldsymbol{\eta}(\mathbf{r})/\partial r_2$$
$$+ \hat{\mathbf{r}}_1 \cdot \partial\boldsymbol{\eta}(\mathbf{r})/\partial r_1 + \frac{\Upsilon(r_1)}{r_1}\frac{\Upsilon(r_2)}{r_2}\hat{\mathbf{r}}_1^\dagger \boldsymbol{\Psi}(\mathbf{r})\hat{\mathbf{r}}_2 + \frac{\Upsilon(r_1)}{r_1}\hat{\mathbf{r}}_1^\dagger \frac{\partial \boldsymbol{\Psi}(\mathbf{r})}{\partial r_2}\hat{\mathbf{r}}_2$$
$$+ \frac{\Upsilon(r_2)}{r_2}\hat{\mathbf{r}}_1^\dagger \frac{\partial \boldsymbol{\Psi}(\mathbf{r})}{\partial r_1}\hat{\mathbf{r}}_2 + \hat{\mathbf{r}}_1^\dagger \frac{\partial^2 \boldsymbol{\Psi}(\mathbf{r})}{\partial r_1 \partial r_2}\hat{\mathbf{r}}_2 + \frac{\Upsilon(r_1)}{r_1}\frac{\Upsilon(r_2)}{r_2}V_0^2 cos\theta_1 cos\theta_2 \qquad (20)$$

where $\Upsilon(r) = 2 + \mathrm{d}ln(\phi(r))/\mathrm{d}ln(r)$.

The tensor $\boldsymbol{\Psi} = \{\Psi_{ij}\}$ is the two-point correlation tensor of the velocity field (Górski 1988),

$$<\mathrm{v_i}(\mathbf{r_1})\,\mathrm{v_j}(\mathbf{r_2})> \equiv \Psi_{ij}(\mathbf{r}) = \Psi_\perp(r)\delta_{i,j} + [\Psi_\|(r) - \Psi_\perp(r)]\hat{r}_i\hat{r}_j \qquad (21)$$

where $\Psi_\|(r)$ and $\Psi_\perp(r)$ are the radial and transverse velocity correlation functions respectively. The spectral representation of these correlation is given by

$$\Psi_{\perp,\|}(r) = \frac{H_0^2 f(\Omega_0)^2}{2\pi^2}\int_0^\infty P(k)K_{\perp,\|}(kr)dk, \qquad (22)$$

where $K_\perp(x) = j_1(x)/x$ and $K_\parallel(x) = j_0 - 2j_1(x)/x$ and $j_l(x)$ is the spherical Bessel function of order $l$.

The vector $\boldsymbol{\eta} = \{\eta_i\}$ is the two-point cross-correlation vector between the density and velocity fields, and it is given by

$$< \mathbf{v}_i(\mathbf{r}_1)\,\delta(\mathbf{r}_2) > \equiv \eta_i(\mathbf{r}) = \frac{H_0 f(\Omega_0)}{2\pi^2} \Xi(r)\hat{\mathbf{r}}_i \qquad (23)$$

where $\Xi(r) = \int_0^\infty kP(k)j_1(kr)dk$. Note that the vector $\eta$ is parallel to $\mathbf{r}$ which is the natural direction in this case.

Numerical calculation of Eq. 20 is rather straightforward. Here we present calculations of this equation for an $8Mpc/h$ Gaussian smoothed $CDM$ model with no selection effect and with $V_0 = 0$. Figs. 1(a-c) show plots of the correlation function in redshift space around a point at a distance of 55, 100 and 500 $Mpc/h$ from a given observer, respectively. Fig. 1d is the correlation function calculated by Kaiser's (1987) formula, $(1 + 2/3f(\Omega_0) + 1/5f^2(\Omega_0))\,\xi(r)$. The Z axis is taken to be the $l.o.s.$ direction. Figs. 1(a-c) demonstrate the different behavior of the correlation function, parallel and perpendicular to $l.o.s.$. They also show the effect of the distance from the observer on the correlation function, which is caused by the $U/r$ term which indicates that it can't be neglected in the analysis of the nearby Universe ($\lesssim 100Mpc/h$).

Fig. 2 shows a linear-log plot of the correlation function in the direction of the $l.o.s.$ and perpendicular to it, calculated from Eq. 20 and compared with the real-space correlation function and with Kaiser's (1987) approximation (angle averaged). The two figures show that the zero crossing of the correlation function in redshift space occurs at a distances shorter than the ones expected from the real space correlation function or from Kaiser's approximation. This can be simply explained by the mass conservation, admitted under the redshift transformation, which amplifies the clustering as $r \to 0$ and therefore attenuates it as $\mathbf{r}_2$ recedes from $\mathbf{r}_1$.

This result is confirmed by the the work of Lilje and Efstathiou (1989) where they used standard CDM simulations to calculate the correlation function along the $l.o.s.$ for galaxies within given transverse separation range (in particular their Figs. 4-7). Fig. 4 of Lilje and Efstathiou shows the N-body calculated correlation function in the real and redshift space along the *l.o.s.* for galaxies with transverse separation in the range of 0.5-1.3 Mpc/h. Note that they ignored the variation of the redshift space correlation function with the distance from the observer. Their Fig. 4 confirms our findings with regard to the *l.o.s.* correlations shown in here in Figs. 1 & 2. Namely the *l.o.s.* redshift-space correlation function is larger than the *l.o.s.* real-space correlation function on small scales and is smaller than it on large scales, where the transition occurs on the scale of 10-20 $Mpc/h$ (for standard CDM). This behavior repeats itself in their Figs. 5-7 where the *l.o.s.* redshift-space correlation function is calculated for a larger transverse separation. Lilje and Efstathiou results also agree with the general behavior found here that the zero crossing scale of the redshift-space function is smaller than that of the real space covariance function. It also disagree with Kaiser's approximation where the ratio of the ratio of the two functions is constant ($\geq 1$).

## V. DISCUSSION

A proper, first order, calculation of the relation between the actual real space density field correlation function and the one observed and measured in redshift space has long be overdue. The formal tools of the the linear theory of gravitational instability could have been easily applied to improve on Kaiser's (1987) approximate results. Given the importance of the redshift-to-real space transformation it is rather surprising that the problem has not been yet properly addressed. The present paper fills up this gap and introduces a straight forward linear calculation of the redshift space covariance functions in the configuration and Fourier spaces.

The present calculation serves a dual purpose. On the one hand this can be used to develop an efficient algorithm for the reconstruction of the real space density field from the observed redshift galaxy distribution. This can be done in the framework of Wiener filtering (Zaroubi *et al.* 1994, Fisher *et al.* 1994 ). On the other hand the redshift space covariance matrix can be used to estimate relevant cosmological parameters by maximum likelihood analysis (*eg.* the power spectrum, $\Omega_0$ and the biasing parameter). Now, the statistical properties of the redshift space perturbation field differ from real space field in two major ways, namely the transformation induces mode-mode coupling and anisotropy of the covariance matrix. The isotropy and orthogonality of the linear regime of the $k$-space covariance matrix is expected in a wide range of cosmological models, and therefore redshift space deviations from these can be used as a probe on $\Omega_0$ or $\beta$, in the case of linear bias. However, in order to make the present calculations into a useful cosmological probe one should complement these by numerical N-body simulations to study the role of non-linear effects and to establish the dynamical range over which the linear theory that has been developed here applies ( Bistolas, Zaroubi and Hoffman, work in progress).

## ACKNOWLEDGMENTS

Stimulating discussions with M. Davis, K. Fisher, O. Lahav, A. Nusser and S. White are gratefully acknowledged. We acknowledge the hospitality of the Institute of Astronomy (Cambridge University), where part of this research was done. YH has been partially supported by The Hebrew University Internal Funds.

**FIGURE CAPTIONS**

Fig. 1: A) The correlation function for the a standard CDM model with 8 $Mpc/h$ smoothing, as measured in the redshift space according to the exact linear calculation around point at 55 $Mpc/h$ distance. Here we take $\phi(r) = 1$ and $V_0 = 0$. All contours are spaced at 0.05 with solid (dashed) lines denoting positive (negative) contours. The heavy solid line denotes the zero line. B) The same as A for point at 100 $Mpc/h$ distance. C) The same as A for point at 500 $Mpc/h$ distance. D) The same correlation function as calculated by Kaiser's (1987) approximation; $\xi_s = (1 + 2/3f + 1/5f^2)\xi_r$.

Fig. 2: The correlation function of the same model described in Fig. 1, parallel to the *l.o.s.* (dotted-dashed line) and perpendicular to it (solid line). The correlation function according to Kaiser's (1987) approximation is also shown (dotted line) and the real space function is given by the dashed line. **Make sure the lines are marked correctly, after my correction!**

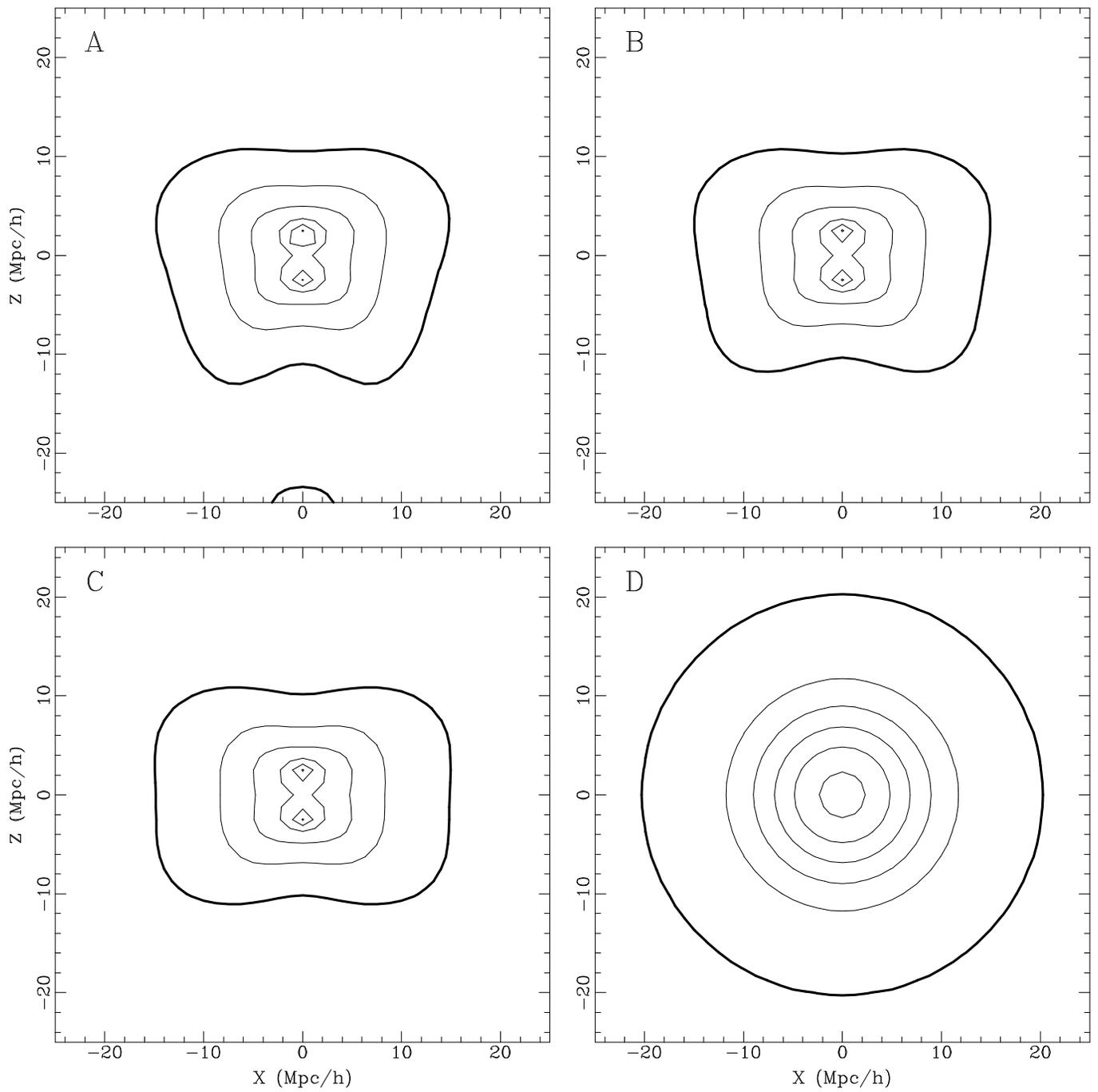

**Figure 1**

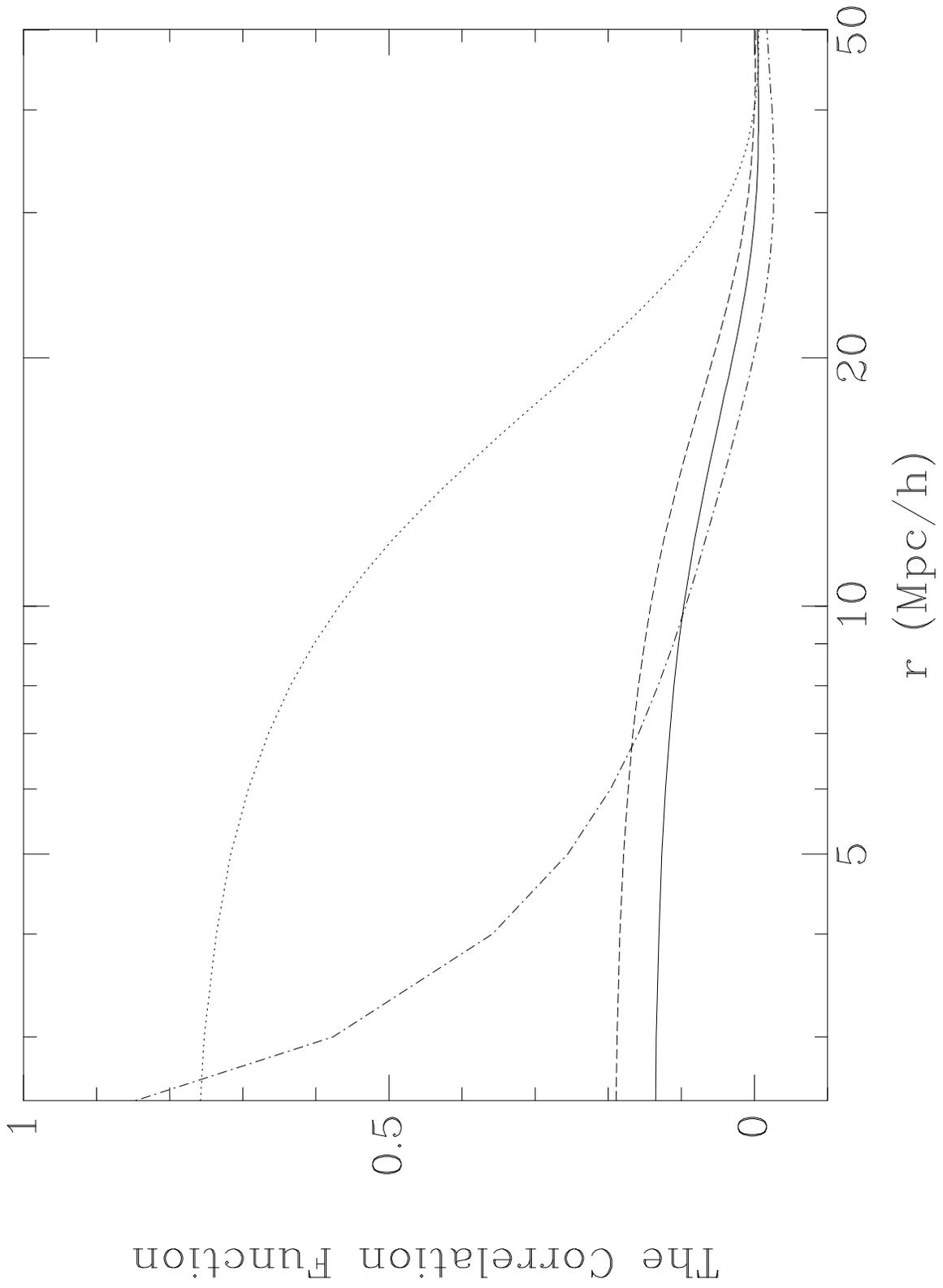

Figure 2